\documentclass[12pt]{iopart}
\usepackage{bm}
\usepackage{graphicx}
\usepackage{iopams}
\expandafter\let\csname equation*\endcsname\relax
\expandafter\let\csname endequation*\endcsname\relax
\usepackage{amsmath}
\usepackage{textcomp, libertine}
\begin{document}
\title[Magnetic field induced polarization enhancement in monolayers of tungsten dichal...]{Magnetic field induced polarization enhancement in monolayers of tungsten dichalcogenides: Effects of temperature}

\author{T. Smole\'nski$^1$, T. Kazimierczuk$^1$, M. Goryca$^1$, M. R. Molas$^2$, K.~Nogajewski$^2$, C. Faugeras$^2$, M. Potemski$^{1,2}$ and P. Kossacki$^1$} 

\address{$^1$  Institute of Experimental Physics, Faculty of Physics, University of Warsaw, ul. Pasteura 5, 02-093 Warsaw, Poland}
\address{$^2$ Laboratoire National des Champs Magn\'etiques Intenses, CNRS-UGA-UPS-INSA-EMFL, 25 rue des Martyrs, 38042 Grenoble, France}

\ead{Tomasz.Smolenski@fuw.edu.pl}
\vspace{10pt}
\begin{indented}
\item[]February 2017
\end{indented}

\begin{abstract}
Optical orientation of localized/bound excitons is shown to be effectively enhanced by the application of magnetic fields as low as 20 mT in monolayer WS$_2$. At low temperatures, the evolution of the polarization degree of different emission lines of monolayer WS$_2$ with increasing magnetic fields is analyzed and compared to similar results obtained on a WSe$_2$ monolayer. We study the temperature dependence of this effect up to $T=60$~K for both materials, focusing on the dynamics of the valley pseudospin relaxation. A rate equation model is used to analyze our data and from the analysis of the width of the polarization deep in magnetic field we conclude that the competition between the dark exciton pseudospin relaxation and the decay of the dark exciton population into the localized states are rather different in these two materials which are representative of the two extreme cases for the ratio of relaxation rate and depolarization rate.
\end{abstract}

\section{Introduction}
The successful isolation of graphene\cite{Novoselov_Science_2004} followed by over a~decade of its intense study has triggered the development of a vast area of research on a variety of two-dimensional (2D) crystals, whose properties substantially differ from those of their bulk counterparts \cite{review-2d-a,review-2d-b,review-2d-c}.
 Among these materials particularly much attention is currently attracted by semiconducting transitional metal
dichalcogenides (s-TMDs) which exhibit  robust optical properties.
The research on these materials is driven by both scientific curiosity and
a prospect of 2D optoelectronic applications.

One of the most intriguing properties of atomically-thin s-TMDs is
the possibility of accessing the valley degree of freedom using circularly
polarized light \cite{Xiao_PRL_2012,Cao_NatCommun_2012,Zeng_NatNano_2012,Mak_NatNano_2012}.  The simplicity of this approach paved the way to
study $T_1$ and $T_2$ parameters of the valley degree of freedom, i.e., the
inter-valley relaxation \cite{Crooker_NatCom_2015} and valley coherence \cite{Jones_NatNano_2013}. Such
experiments are typically carried out at cryogenic temperatures, which
allows one to minimize the influence of thermal effects obscuring the
investigated phenomena.

On the other hand, the bulk of application-oriented studies on s-TMD monolayers is
carried out at room temperature. Due to extremely high exciton binding
energy (of hundreds of meV \cite{Heather_NanoLett_2015,Ye_Nature_2014}) and short radiative lifetime (in the range of
picoseconds \cite{Lagarde_PRL_2014}), the optical properties of s-TMD monolayers remain robust under such
conditions, as evidenced by efficient room-temperature photoluminescence
(PL) \cite{Amani_Science_2015}.
A disparity between these two regimes immediately raises a question about the
changes in the intermediate temperature range, which is of particular interest, since even as
basic quantity as the PL intensity exhibits a non-monotonic dependence on 
temperature \cite{Lezama_NanoLett_2015}.

\begin{figure}
\centering
\includegraphics[height=90mm]{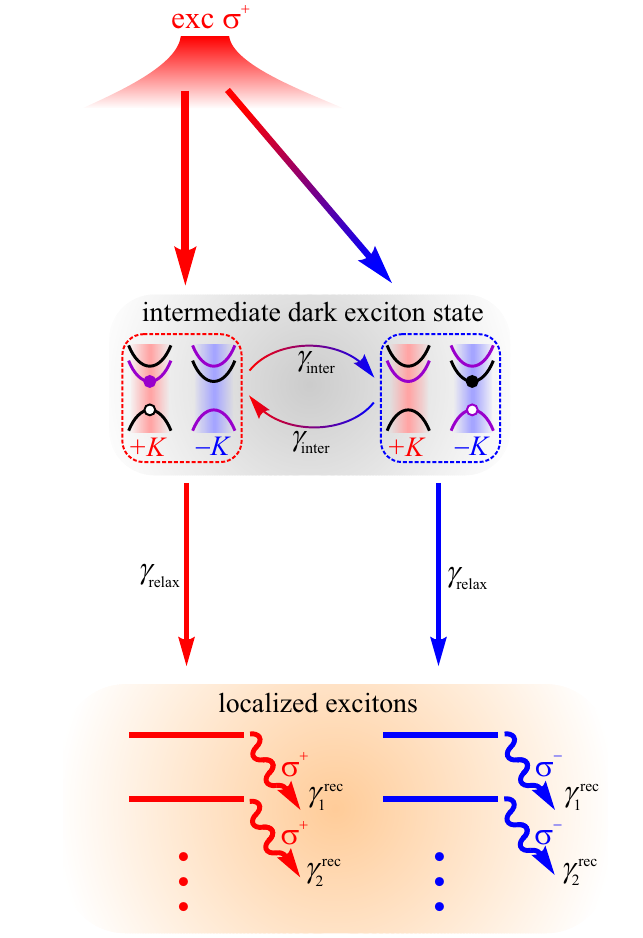}
\caption{Relaxation pathways leading to the formation of localized excitons via dark excited states
according to Ref. \cite{Smolenski_PRX_2016}. A variation of the inter-valley relaxation rate $\gamma_\textrm{inter}$ with the magnetic field is responsible for the dip in the field dependence of the localized exciton optical orientation efficiency. After Ref. \cite{Smolenski_PRX_2016}. \label{schemat}}
\end{figure}

In this letter we present a study of temperature dependence of the magnetic-field-induced polarization enhancement (FIPE) of the optical orientation of localized excitons (LEs) in WS$_2$ and WSe$_2$ monolayers.
Earlier investigations \cite{Smolenski_PRX_2016} established a link between the occurence of the FIPE for monolayer WSe$_2$ and the inter-valley relaxation of dark excitons, which could precede the localized excitons in the relaxation path (see Fig. \ref{schemat}). The key element in the proposed model was the process of inter-valley relaxation of the dark excitons, which could be effectively suppressed by the magnetic field. Regardless of its rate, the inter-valley relaxation is stopped by localization of the dark excitons (conversion to the population of localized excitons). Thus, the analysis of the polarization degree of the subsequent localized exciton recombination shed some light on the nature of inter-valley relaxation of dark excitons, which is much more elusive than the analogous relaxation of bright excitons \cite{Yu_PRB_2014}.

\section{Methods}
The experiments described in this paper were performed on 2 sets of samples: monolayers of WS$_2$ and WSe$_2$. In each case the monolayer flakes were obtained by mechanical exfoliation of bulk crystals involving the use of a chemically clean back grinding tape and a polydimethylsiloxane stamp.
The exfoliated flakes were non-deterministically transferred onto Si substrates covered with either 90 nm thick (WSe$_2$) or 300 nm thick (WS$_2$) layer of thermally grown SiO$_2$. Prior to that step the substrates were ashed with oxygen plasma to rid their top surfaces of possible organic contaminants. The monolayers of interest were first identified by their distinctive optical contrast and then double checked with photoluminescence and Raman scattering measurements.

Our optical investigations were based on photoluminescence spectroscopy. The samples were excited using either 561~nm or 647~nm continuous-wave diode lasers. The emitted light was analyzed using a~0.5~m spectrometer equipped with a CCD camera. The polarization state of the light was controlled by means of a~combination of Glan-Taylor polarizers and achromatic $\lambda/4$ waveplates incorporated in the excitation as well as in the detection path. 

During the experiments, the samples were placed in a variable temperature insert (VTI) of a helium cryostat mounted inside a superconducting magnet. All measurements were carried out in the Faraday geometry.

\section{Results}
\subsection{Field-dependent optical orientation in WS$_2$ monolayer}
So far the FIPE effect has been only observed in WSe$_2$ monolayers \cite{Smolenski_PRX_2016}. In the reported experiments
a rather weak magnetic field of $~20$ mT was shown to significantly enhance the efficiency of the optical orientation, $\eta(B)$, of the localized excitons, leading to a~dip-like feature in the $\eta(B)$ dependence at $B=0$. This observation was interpreted as related to the field-dependent inter-valley relaxation of the dark excitons. The fact that the energetically lowest excitonic state is optically dark in WSe$_2$ monolayers \cite{Kosmider_PRB_2013,Echeverry_PRB_2016,Molas_2DMat_2017} has been a crucial point in the interpretation of the observation of the FIPE. Monolayer WS$_2$  is another system with an optically inactive exciton ground state, and as such might be expected to exhibit the FIPE too.

\begin{figure*}
\centering
\includegraphics[width=155mm]{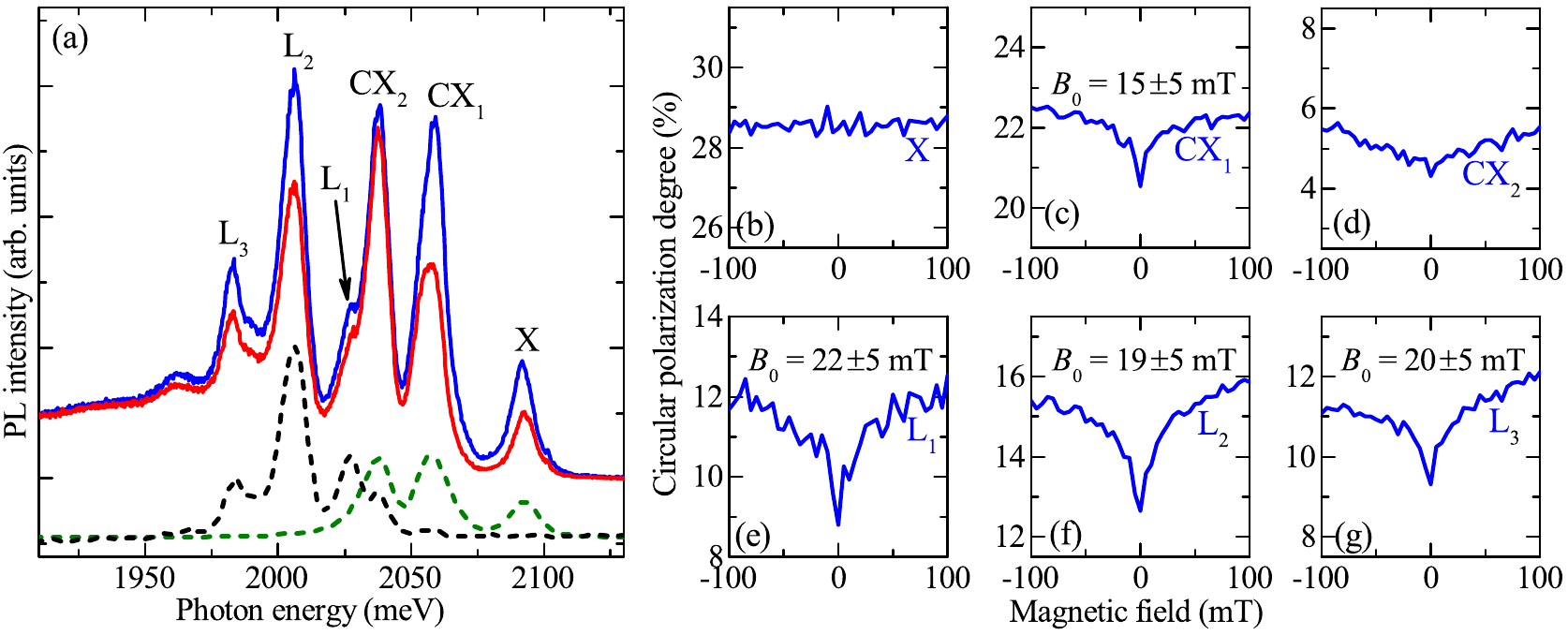}
\caption{(a) PL spectra of a WS$_2$ monolayer in the co- (blue line) and cross-circular (red line) polarization.
Dashed lines show two time-gated spectra: one measured during 30~ps shortly after the excitation pulse (green line) and one measured during the time window of 850~ps staring 200~ps after the excitation pulse (black line). (b-g) Field dependence of the efficiency of optical orientation of each of the emission lines indicated in panel (a). \label{fig1}}
\end{figure*}

Figure \ref{fig1} presents the effect of optical orientation of excitons in a~WS$_2$ monolayer at low temperature ($T=1.5$~K). The optical orientation manifests itself as a difference between the PL spectra measured in two circular polarizations upon circularly polarized excitation, as seen in Fig. \ref{fig1}(a). Such a difference is observed with different strength for various emission lines, which correspond to the recombination processes involving either free or localized excitons. Both kinds of excitons can be distinguished on the basis of their lifetime, as free excitons decay in single picoseconds while localized excitons can live for up to a few nanoseconds \cite{Smolenski_PRX_2016,Wang_PRB_2014}. In our case, the according time-resolved data were obtained in an auxiliary experiment using a streak camera and pulsed excitation of the PL, which allowed us to obtain  reference time-gated spectra plotted with green and black dashed lines in Fig. \ref{fig1}(a). The signal drawn with a black curve was integrated in a window from 200~ps to 1050~ps after the laser pulse, i.e., in the time range after the recombination of free excitons when only the localized excitons contributed to the spectrum. Using this method we unequivocally determined that the lines denoted in Fig. \ref{fig1}(a) by X and CX$_1$ are related the the free excitons (neutral and charged, respectively). Due to its intermediate lifetime, the line CX$_2$ cannot be easily classified as arising from either free or localized excitons. Based on the results of reflectivity experiments on similar samples\cite{jadczak_arxiv_2016}, we tentatively ascribe it to another type of a charged exciton. Still, in the following analysis for the WS$_2$ monolayer we will limit ourselves to the three lowest-energy lines (L$_1$-L$_3$) in the spectrum, which can be unambiguously attributed to the localized excitons.

Similarly to the approach used for monolayer WSe$_2$ \cite{Smolenski_PRX_2016}, we have repeated the measurements of optical orientation for various magnetic fields, with the results shown in Fig. \ref{fig1}(b-f). As expected, each of the lines recognized as originating from the localized states exhibits the effect of robust FIPE, i.e., a~significant variation of the optical orientation efficiency at a~rather weak magnetic field of 20--22~mT. Upon application of the magnetic field the optically-induced polarization is clearly enhanced, which may be interpreted as a signature of the slowdown of inter-valley relaxation in the dark exciton population\cite{Smolenski_PRX_2016}.

Interestingly, the FIPE is also observed for the line denoted in Fig. \ref{fig1}(a) as CX$_1$, most probably related to the free charged excitons \cite{Mak_NatMater_2013,Plechinger_PSSL_2015}. In its present form, our model does not automatically account for this effect, which requires a special treatment, e.g., the assumption that dark excitons contribute at least partially to the formation of the trions. A verification of this assumption would necessitate precise polarization- and time-resolved measurements on gated structures, which remain beyond the scope of this work.

A separate issue is the relative amplitude of the FIPE (Fig. \ref{fig1}(d-f)). Depending on the particular localized exciton line, the amplitude of this effect in WS$_2$ monolayer is equal to about 2 percentage points. It is significantly weaker than the effect observed for monolayer WSe$_2$, which is about 25 percentage points \cite{Smolenski_PRX_2016}.
Such a disparity cannot be explained by differences in the overall optical orientation efficiency, which was quite similar under our experimental conditions ($\approx 13$\% for monolayer WS$_2$ and $\approx 30$\% for monolayer WSe$_2$ at $B=100$~mT, respectively).
We can speculate that the discussed disparity reflects a different interplay between the exciton localization time and the rate of inter-valley relaxation of the itinerant excitons. In particular, already at $B=0$ the localization time in a~WS$_2$ monolayer might be short compared to the inter-valley relaxation time. Under such conditions, an additional extension of the relaxation time by the magnetic field would have a limited effect, as in the case of the experiment. This conjecture will be revisited when discussing the temperature dependence of the width of the dip in the field dependence of the orientation efficiency.

\begin{figure*}
\includegraphics[width=155mm]{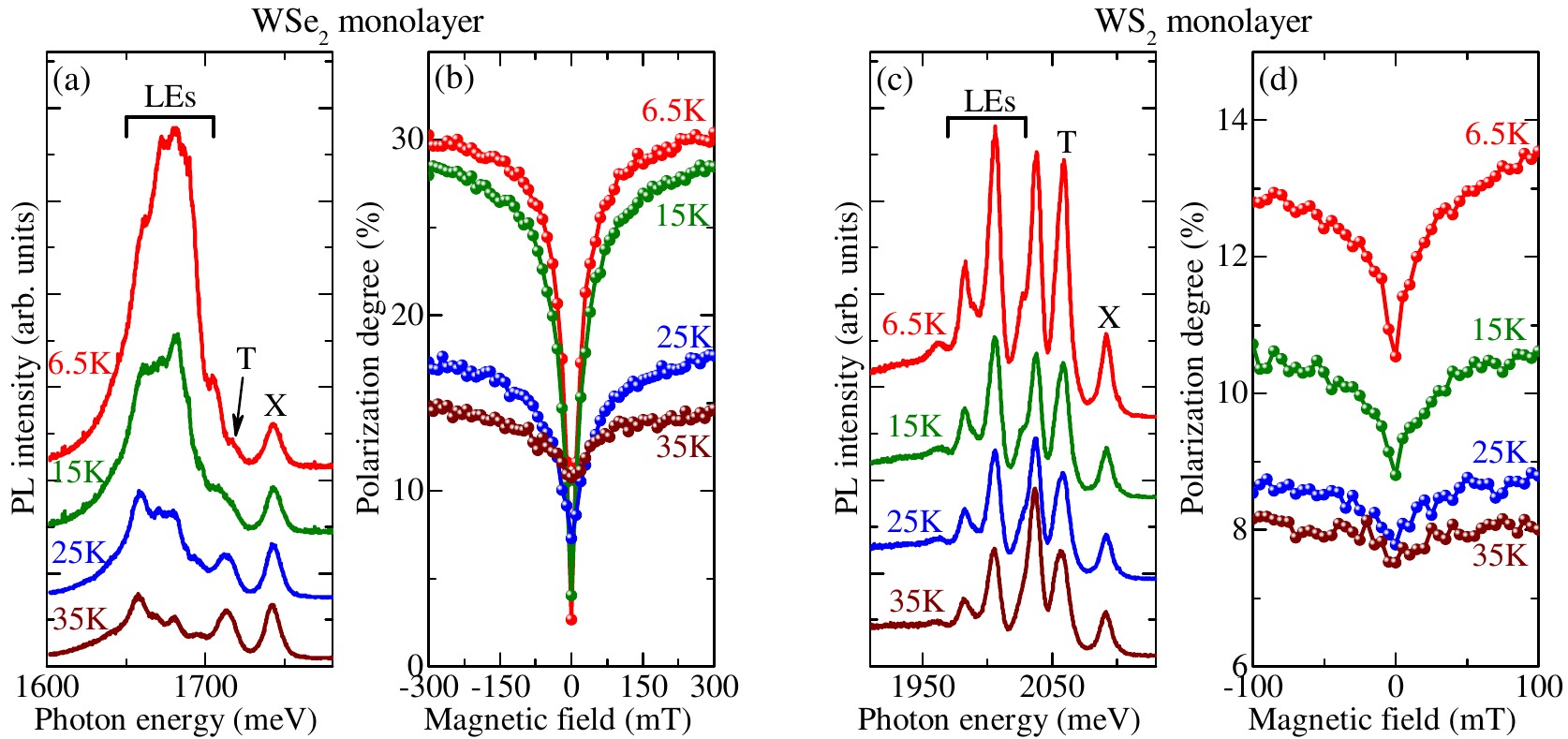}
\caption{Representative results illustrating the temperature dependence of the optical properties of monolayers of WSe$_2$ (a,b) and WS$_2$ (c,d). Panels (a) and (c) show the unpolarized PL spectra measured at $B=0$ with vertical offset added for clarity. Panels (b) and (d) demonstrate the field dependence of the optical orientation of the localized exciton emission in the range marked in the respective PL spectrum. No offset was added to the data presented in panels (b) and (d).  \label{fig2}}
\end{figure*}

\subsection{Temperature dependence}

Having established that monolayer WS$_2$  also exhibits the FIPE effect similar to that displayed by monolayer WSe$_2$, we repeated measurements for both systems at different bath temperatures. An overview of the observed changes is presented in Fig. \ref{fig2}. In both materials the increase in temperature leads to a~systematic reduction of the PL intensity of the LE lines, which is interpreted in terms of thermal activation (de-localization) of the LEs \cite{Godde_PRB_2016}. This quenching effect limits the range  of experimentally feasible temperatures to about 60~K, particularly for monolayer WSe$_2$. Within this limit we were able to accurately determine the efficiency of optical orientation of the LE and to study its field dependence. 

\begin{figure}
\centering
\includegraphics[width=85mm]{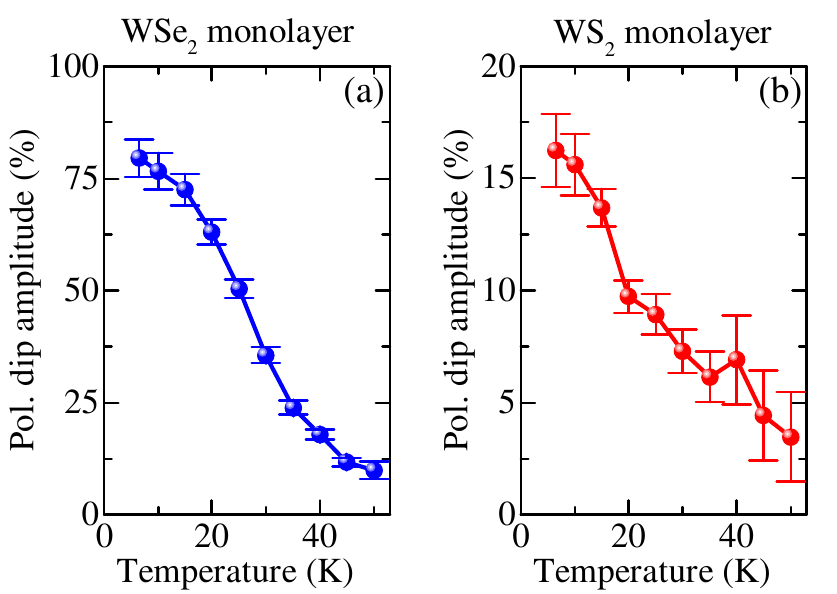}
\caption{Temperature dependence of the relative amplitude of the dip in the field-dependence of the optical orientation of (a) WSe$_2$ monolayer and (b) WS$_2$ monolayer. \label{fig3}}
\end{figure}

As shown in Fig. \ref{fig2}(b) and  \ref{fig2}(d), the characteristic dip in  $\eta(B)$ is present also at elevated temperatures. Still, its amplitude is gradually reduced upon increasing the temperature. We note that at the same time the overall optical orientation efficiency (defined by the level at $B>100$~mT) also decreases. Yet,
the disappearance of the dip is more rapid. We visualize it in Fig. \ref{fig3}, which presents the temperature dependence of the relative amplitude of the dip, i.e., the amplitude of the dip divided by the orientation efficiency at $B>100$~mT. The quenching of the dip amplitude is similar in both analyzed materials. Notably, the reduction of the dip to half of its initial relative amplitude occurs at comparable temperature (about 25~K for monolayer WSe$_2$  vs. 20~K for monolayer WS$_2$).

\begin{figure}
\centering
\includegraphics[width=85mm]{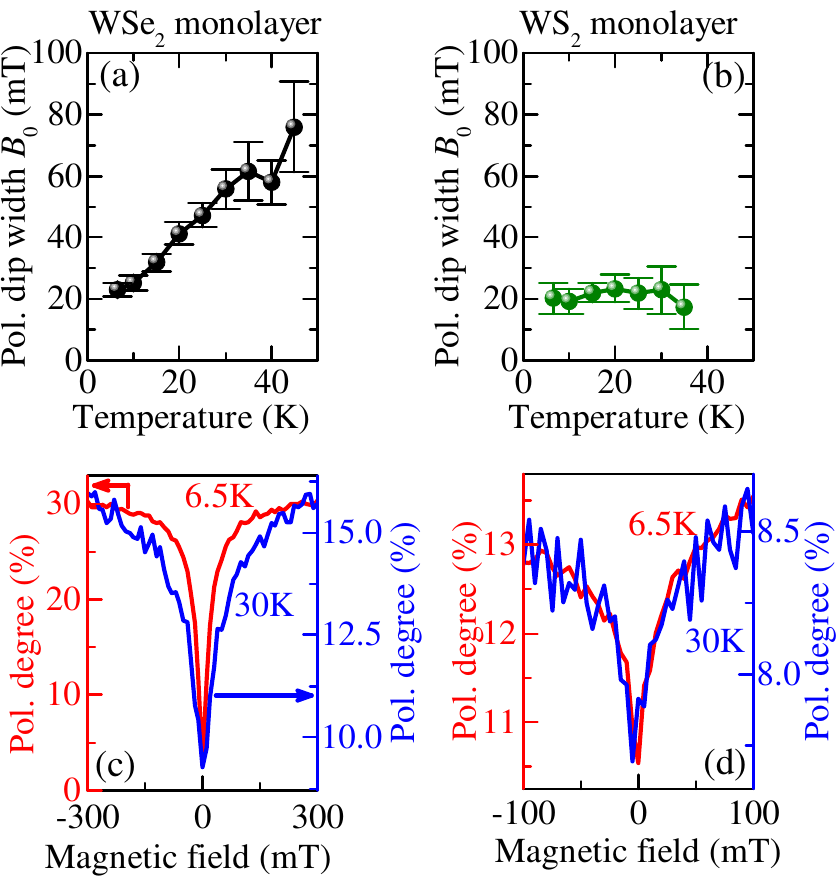}
\caption{Temperature evolution of the HWHM of the dip in the field-dependence of the optical orientation of (a) WSe$_2$ monolayer and (b) WS$_2$ monolayer. Panels (c) and (d) show a direct comparison between the shapes of the profile at 6.5~K and 30~K for WSe$_2$ and WS$_2$, respectively.   \label{fig4}}
\end{figure}

The main complication associated with an analysis of the FIPE amplitude is a~possible opening of other relaxation channels for the dark excitons when increasing the temperature. Much more reliable is the second characteristic of the  dip observed in the optical orientation efficiency, namely its width. Temperature dependence of the extracted width (half width at half maximum, HWHM) is plotted in Fig. \ref{fig4}(a,b) and evidences a~qualitatively different behavior of WSe$_2$ and WS$_2$ monolayers. Within the experimental accuracy the latter system exhibits a~constant width of the dip in the $\eta(B)$ dependence, while for  WSe$_2$ the dip becomes clearly broader upon increasing the temperature. This observation can be  confirmed directly by comparing the magnetic field dependence of the optical orientation efficiency at low (6.5~K) and elevated (30~K) temperature, which is shown in Fig. \ref{fig4}(c,d). 

Our interpretation of the observed behavior is related to the temperature variation of the underlying relaxation processes. In general, the width of the dip in the optical orientation efficiency reflects the magnetic field which slows down the inter-valley relaxation of the dark excitons to the time scales of the competing localization process. Both of these rates depend on temperature, but in the studied range we expect a~significant variation of the inter-valley relaxation rate. This expectation is based on the fact that the exchange interaction behind the inter-valley relaxation scales as the square of the wave vector, $k^2$ \cite{Glazov_PSSB_2015}, and  is therefore proportional to the temperature of the excitons. If the rate of the localization process is less dependent on the temperature (e.g., because of low phonon population in the relevant temperature range) then the width of the dip should increase accordingly, as higher and higher magnetic field is required to counteract the increase in the inter-valley relaxation rate.

The scenario described above is fully consistent with the data obtained for monolayer WSe$_2$  (Fig. \ref{fig4}(a)), but does not agree with the results for monolayer WS$_2$ (Fig. \ref{fig4}(b)), despite that a~similar temperature dependence of the dark exciton inter-valley relaxation rate is expected in both cases. A~possible explanation of this inconsistency is different relation between the rates of the discussed processes. If the localization rate outpaces the inter-valley relaxation rate even at $B=0$ then the width of the dip will not correspond to slowing down of the latter one to the value of the former one. Instead, the width of the dip would correspond to the magnetic field which reduces the inter-valley relaxation to a given fraction (say, by factor of 2 at HWHM), independent of the threshold value of the localization rate.

The presented reasoning can be illustrated in a simple way with the following  equations. In the model sketched in Fig. \ref{schemat} the valley polarization degree of the localized exciton varies with time as 
\begin{equation}
\eta_\mathrm{DE} = \eta_{\infty} e^{-\gamma_\mathrm{inter} t}.
\end{equation}
However, the quantity relevant for the experiment is the final (i.e., after the localization but before the recombination) polarization degree of the localized excitons, which can be obtained by integrating separately each of the dark exciton populations with a decay rate $\gamma_\mathrm{relax}$. In that way we obtain the expected degree of polarization of the photoluminescence:
\begin{equation}
\eta = \frac{\eta_{\infty}}{1+ \gamma_\mathrm{inter}/\gamma_\mathrm{relax} }.
\end{equation}
Assuming that $\gamma_\mathrm{inter}$ is the only variable parameter, we can derive the relation between its value at HWHM of the dip and at $B=0$:
\begin{align}
\eta_{1/2} & = \frac{1}{2}\left(\eta_0 + \eta_\infty \right) \\
\frac{\eta_{\infty}}{1+ \gamma_\mathrm{inter,1/2}/\gamma_\mathrm{relax}} 
& = \frac{1}{2} \left( \frac{\eta_{\infty}}{1+ \gamma_\mathrm{inter,0}/\gamma_\mathrm{relax}}
+\eta_\infty \right).
\end{align}
and therefore:
\begin{equation}
\gamma_\mathrm{inter,1/2} = \frac{ \gamma_\mathrm{inter,0} } {2 + \gamma_\mathrm{inter,0} / \gamma_\mathrm{relax} }. \label{eq:hwhm}
\end{equation}
Eq. \ref{eq:hwhm} clearly shows the two regimes. If $\gamma_\mathrm{inter,0} \gg \gamma_\mathrm{relax}$ (case more suitable for monolayer WSe$_2$) then $\gamma_\mathrm{inter,1/2} \approx \gamma_\mathrm{relax}$. The magnetic field needed to supress growing $\gamma_\mathrm{inter,0}$ to the threshold of $\gamma_\mathrm{relax}$ is larger and larger, hence the increasing dependence in Fig. \ref{fig4}(a). Conversely, if $\gamma_\mathrm{inter,0} \ll \gamma_\mathrm{relax}$ (case applicable to monolayer WS$_2$) then $\gamma_\mathrm{inter,1/2} \approx \gamma_\mathrm{inter,0}/2$. Standard mechanisms of Dyakonov-Perel type predict that
$\gamma_\mathrm{inter}(B)= \gamma_\mathrm{inter,0} f(B)$, where the shape function $f(x)$
is not related to the $\gamma_\mathrm{inter,0}$ value itself. In such a case, the magnetic field required to reduce $\gamma_\mathrm{inter}$ to the half of its initial value is constant (independent of $\gamma_\mathrm{inter,0}$), which explains the behavior illustrated in Fig. \ref{fig4}(b). 

In our considerations we neglected a number of additional effects, such as the influence of other relaxation pathways (e.g., including the bright excitons) or temperature dependence of the localization or de-localization rate. Similarly, our working assumptions of $\gamma_\mathrm{inter,0} \gg \gamma_\mathrm{relax}$ or $\gamma_\mathrm{inter,0} \ll \gamma_\mathrm{relax}$ should be regarded as serving illustrative purposes only, since the actual difference between these rates might be less pronounced.
Still, on the qualitative level our model satisfactorily explains the experimental data suggesting that the decay of the dark exciton population is more rapid in monolayer WS$_2$ than in monolayer WSe$_2$.

\section{Summary}
In summary, we have shown that the application of a small magnetic field significantly enhances the efficiency of optical orientation not only in WSe$_2$ but also in WS$_2$ monolayers. Thus, we confirm our original hypothesis that the B-enhanced optical orientation is characteristic of s-TMD monolayers in which the energetically lowest  excitonic state is dark (optically inactive in the first approximation). The amplitude of this enhancement is observed to be suppressed by temperature for both WSe$_2$ and WS$_2$ monolayers. In particular, the B-enhanced optical orientation becomes hardly detectable at temperatures above 50K. The characteristic range of magnetic fields where the enhancement is observed is however found to expand with temperature in monolayer WSe$_2$  whereas is independent of temperature in monolayer WS$_2$. This distinct behavior is interpreted in terms of different regimes for the respective rates for intervalley relaxation and decay of dark excitons. We anticipate that the intervalley relaxation of dark excitons is fast as compared to their decay in monolayer WSe$_2$ while the opposite relation holds for monolayer WS$_2$. 
Our observation undoubtedly requires further investigation that will provide it with not only phenomenological but also theoretical explanation.

\ack
The work was supported by National Science Center, Poland under project no. DEC-2015/17/B/ST3/01219, the European Research Council (MOMB project no. 320590), the European Graphene Flagship,
and the ATOMOPTO project carried out within the TEAM programme of the Foundation for Polish Science co-financed by the European Union under the European Regional Development Fund. One of us (T.S.) was supported by
the Polish National Science Centre through PhD scholarship grant
DEC-2016/20/T/ST3/00028.

\section*{References}
\bibliographystyle{iopart-num}

\providecommand{\newblock}{}

\end{document}